\documentclass[sigconf, nonacm]{acmart}

\usepackage{booktabs}
\usepackage{subfigure}
\usepackage{adjustbox}
\usepackage{newfloat}

\begin{document}

\title{Teaching Scrum with a focus on compliance assessment}
\author{Marco Torchiano} \orcid{0000-0001-5328-368X} 

\email{marco.torchiano@polito.it}
\affiliation{%
  \institution{Politecnico di Torino}
  \streetaddress{C.so Duca degli Abruzzi 24}
  \city{Torino}
  \country{Italy}
  \postcode{10129}
}

\author{Antonio Vetrò} \orcid{0000-0003-2027-3308} 
\email{antonio.vetro@polito.it}
\affiliation{%
  \institution{Politecnico di Torino}
  \streetaddress{C.so Duca degli Abruzzi 24}
  \city{Torino}
  \country{Italy}
  \postcode{10129}
}

\author{Riccardo Coppola} \orcid{https://orcid.org/0000-0003-4601-7425}
\email{riccardo.coppola@polito.it}
\affiliation{%
  \institution{Politecnico di Torino}
  \streetaddress{C.so Duca degli Abruzzi 24}
  \city{Torino}
  \country{Italy}
  \postcode{10129}
}

\renewcommand{\shortauthors}{Torchiano et al.}

\begin{abstract}
The Scrum framework has gained widespread adoption in the industry for its emphasis on collaboration and continuous improvement. However, it has not reached a similar relevance in Software Engineering (SE) curricula. This work reports the experience of five editions of a SE course within an M.Sc. Degree in Computer Engineering.  The course primary educational objective is to provide students with the skills to manage software development projects with Scrum. The course is based on the execution of a team project and on the definition of qualitative and quantitative means of assessment of the application of Scrum. The conduction of five editions of the course allowed us to identify several lessons learned about time budgeting and team compositions in agile student projects and its evidence of the applicability of the framework to software development courses.
\end{abstract}





\maketitle

\section{Introduction}

Scrum is an agile framework primarily used in software development management, characterized by its iterative approach and emphasis on collaboration, adaptability, and continuous improvement. In current practice, Scrum has gained widespread recognition for its ability to enhance productivity, efficiency, and transparency in various tasks, including -- but not limited to -- software development. Its importance in university education lies in its capacity to cultivate essential skills such as teamwork, communication, problem-solving and critical thinking, preparing students for the dynamic and rapidly evolving demands of the modern workforce. 

Recent evidence from scientific literature underlines that Scrum is not always considered a cornerstone for modern software development by Software Engineering curricula and that in many cases reference
curricula do not consider Scrum with a great relevance in
knowledge areas nor in the time scheduled for it \cite{perez2018improving}.

However, several pieces of research in Software Engineering education provided successful examples of integrating the teaching of Scrum fundamentals in development courses. Experience reports about the application of Scrum in developing a project in an undergraduate Software Engineering capstone course underline a positive appreciation by the involved students \cite{mahnic2011capstone}, and the possibility of obtaining student projects that can meet real-world quality standards and be adopted for use by Universities after the termination of the course \cite{ding2017case}. The construction of a development setup using Scrum also allows courses to involve IT professionals, allowing students to gain a better appreciation of the inherent challenges involved in crafting larger and more realistic software applications than traditional, small-scale course projects \cite{linos2020involving}. 

This manuscript reports the experience of introducing the teaching of the Scrum framework in a Master's level Software Engineering course and describes an assessment approach, both qualitative and quantitative, to evaluate the outcomes of the students' development activities as well as the coordination and collaboration within teams. We detail the organization of the course and the results of its first five editions. On this basis, we discuss the educational results obtained and the key lessons learned.

\section{Course organization}

In this section we describe the way the course is organized, emphasizing the educational goals, the outline of the course, the way the teams were composed, and the project(s) assigned to the students.

\subsection{Educational goals}
The primary educational objective of the course is to provide the students with the essential skills and knowledge required to effectively manage and execute software projects using agile methodologies. Our focus is on instilling a comprehensive understanding of the agile development approach, emphasizing hands-on experience in carrying out team-based software projects efficiently and in an inclusive manner, i.e. enhancing the peculiarities of each person. 

In pursuit of this educational goal, we adopt the Scrum framework as a reference agile methodology: Scrum, in contrast to other agile methodologies like Kanban, provides a well-structured and systematic framework that is particularly suitable for those who are new to agile practices. Scrum's emphasis on iterative development facilitates a smoother learning curve for participants, allowing them to grasp the fundamental concepts of agile project management in a more structured manner. Scums’s regular checkpoints are helpful to receive continuous feedback and adapt to teachers’ indications, allowing students to conduct a reflective practice led by the critical analysis of their work. In addition, the focus on collaborative teamwork reduces students' dropouts and increases the overall achievement of learning outcomes. 

Although the main object of the course is the process and not the technology, we devote the majority of time to hands-on experience for delivering working software: this is crucial for developing the skills necessary to handle the complexity of software project management in times of dynamic environments and volatile requirements. In summary, at the end of the course, the students acquire basic knowledge of:
\begin{itemize}
\item the principles of agile development and how they are implemented in Scrum; 
\item how to manage a team-based software project, planning activities and adapting to changes during its evolution;
\item how to critically analyse the processes involved in software development and improve them;
\item how to set up self-organizing teams and arrange teamwork combining efficiency and inclusiveness; 
\item how to manage stakeholders' collaboration and report on project's achievements and fulfilment of requirements;  
\item how to manage and improve software quality.
\end{itemize}


\subsection{Course outline}
The course lasts 14 weeks and it is conceptually divided into two phases: a \textit{how-to-do} phase (6 weeks) and a \textit{do} phase (8 weeks). Each week includes 4,5 hours of lecture/activity, three of which are held in parallel rooms: in total, each student attends 60 hours of lecture, while the overall teaching effort is about 150 hours.

In the \textit{how-to-do} phase, we introduce the theoretical principles of Agile and we show how a typical Scrum development process is organized. We carry on a pilot project through a series of workshops focused on the key activities in the Scrum method, each corresponding to a workshop assigned to the students:

\begin{itemize}
    \item W1 - User stories 
    \item W2 - Estimation 
    \item W3 - Planning
    \item W4 - Daily scrum
    \item W5 - Review
    \item W6 - Retrospective
\end{itemize}

Each workshop is introduced by a review of the corresponding Scrum activity and related artefacts (e.g., the Scrum board) and instructions on how to perform the activity. Then each team performs the activity in class: during the activity, they receive feedback and guidance from the three instructors. At the end, teams gather again in a plenary session where instructors summarize the most frequent errors and difficulties encountered, and how to face them.

In the \textit{do} phase the students carry on the real project (see Sec. \ref{sec:project}). In this phase, in addition to the project's related activities, additional theoretical lectures are given on a variety of ancillary topics: agile contracts, technical debt (TD) and software quality measurement, other agile approaches, and free software licenses. Practical activities are also conducted in conjunction with some of the lectures, for example on managing TD, choosing a proper free software license or making an Elephant Carpaccio exercise. 

\subsection{Team composition}
We form teams right after the first week. As inclusiveness is a fundamental goal in our course, we strive to achieve homogeneity in teams by balancing a variety of aspects that are assessed through an online form: provenance and nationality; gender; working student; knowledge of basic programming languages (Java, Python, SQL, Cpp), knowledge of web development languages and techniques (e.g., HTML, JS, Restful APIs, Docker, etc.); previous knowledge on software engineering concepts and tools (e.g., UML, testing, code reviews, GIT), previous knowledge on agile. The goal is to avoid the isolation of the less technically skilled, non Italian students, etc.

Through the years, the total number of teams ranged from a minimum of 11 in the first edition (A.Y. 2019/20, 62 students) to a maximum of 20 in the A.Y. 2022/23 (122 students). In the last edition (i.e., A.Y. 2023/24) there were 119 students organized in 18 teams. Attendance to workshops and to the project's reporting appointments (see Sec. \ref{sec:project}) is mandatory, we allow only four absences.

\subsection{The project assignment}
\label{sec:project}
During the \textit{do} phase of the course, the students carry on four two-week sprints on the real project. In the different editions of the course, the students have been tasked with the full-stack development of web applications, with varying domains and requirements throughout the years, e.g. a thesis management system for a university, a hike tracker application, or a solidarity purchasing group. During the project, one of the teachers acts as the Product Owner and answers the questions on a Telegram chat so that every team is informed of the feedback. Another teacher plays the role of Scrum Master and is available to teams for counselling on how to apply Scrum or to discuss and help resolve collaboration issues. Each person must devote 16 hours of work during a sprint.

At the end of every sprint, teams carry on a product review and a sprint retrospective. 
In product reviews, teams present in front of one instructor and another team (both randomly assigned at each sprint) which plays the role of stakeholders: after the 20-minute demo of the working software, they provide feedback as a pair of positive aspects (e.g., high usability, a well working functionality, etc.) and a negative aspect (e.g., a bug noticed during the demo, or a user story not well developed, etc.), the latter of which must be reported in the GitHub repository and properly addressed in the next sprint. The teams then swap: stakeholders become presenters and vice-versa. During both reviews, the instructor makes questions to assess the correct development of user stories and other aspects and takes notes to be subsequently analyzed (see Sec. \ref{sec:assessment}). Notice that the definition of done adopted in the course includes the following aspects: unit tests passed; code reviewed performed; code on the version control system; E2E tests performed. Teams might add additional elements at their will.

Concerning the sprint retrospectives, teams have 20 minutes to summarize the content of the \textit{retrospective report} and to answer questions from the teacher (randomly assigned at every sprint). The retrospective report is a template\footnote{It can be accessed at \url{https://anon.to/Ld2jus}} to be filled with a mix of quantitative and qualitative data, on the following aspects: stories developed, effort and overall estimation error; tests performed and related effort; technical debt effort and achievements (sprint 3 and 4 only); reflections on estimation errors and their causes, coordination issues, improvements achieved and planned for the next iteration. 

Students can use technologies of their choice, both for development and for supporting team coordination (e.g., many use Discord channels to make Scrum meetings). However, they are required to include the following technologies/tools in their development process: YouTrack for tracking tasks, effort and project progress; GitHub as hosting repository; SonarCloud to manage TD; Docker to containerize the application in two releases (at sprint 2 and at sprint 4). All repositories (GitHub, SonarCloud, YouTrack, Docker Hub) need to be public for assessment purposes and for encouraging collaboration even among teams. At the end of the project, teams must upload a 7-minute teaser video on YouTube which shows how the application performs a provided scenario. The students can use part of the time also to show the strengths of their application.

\section{Assessment method}
\label{sec:assessment}

The grading is performed by assessing both team and individual performance. Team assessment weights 80\%, while individual assessment weights 20\%.
The individual assessment entails both theoretical knowledge and contribution provided to teamwork, especially during the demos and retrospectives: notes taken by instructors during sprint reviews and retrospectives contribute to this part of the evaluation.

The team assessment is performed using the following criteria:

\begin{itemize}
    \item \emph{Team coordination and improvement}, which includes the proper use of the project tracking tool (YouTrack), planning and time management, and improvement of corresponding tracked metrics  (see Sec.  \ref{sec:team}) during the project;
    \item \emph{Quality} of sprint reviews (demos) and retrospectives meetings;
    \item \emph{Doneness}, which means testing of product, management of technical debt, and deployed releases. 
\end{itemize}

\subsection{Assessment of Teams}
\label{sec:team}
\begin{figure*}
    \centering
    \includegraphics[width=1\linewidth]{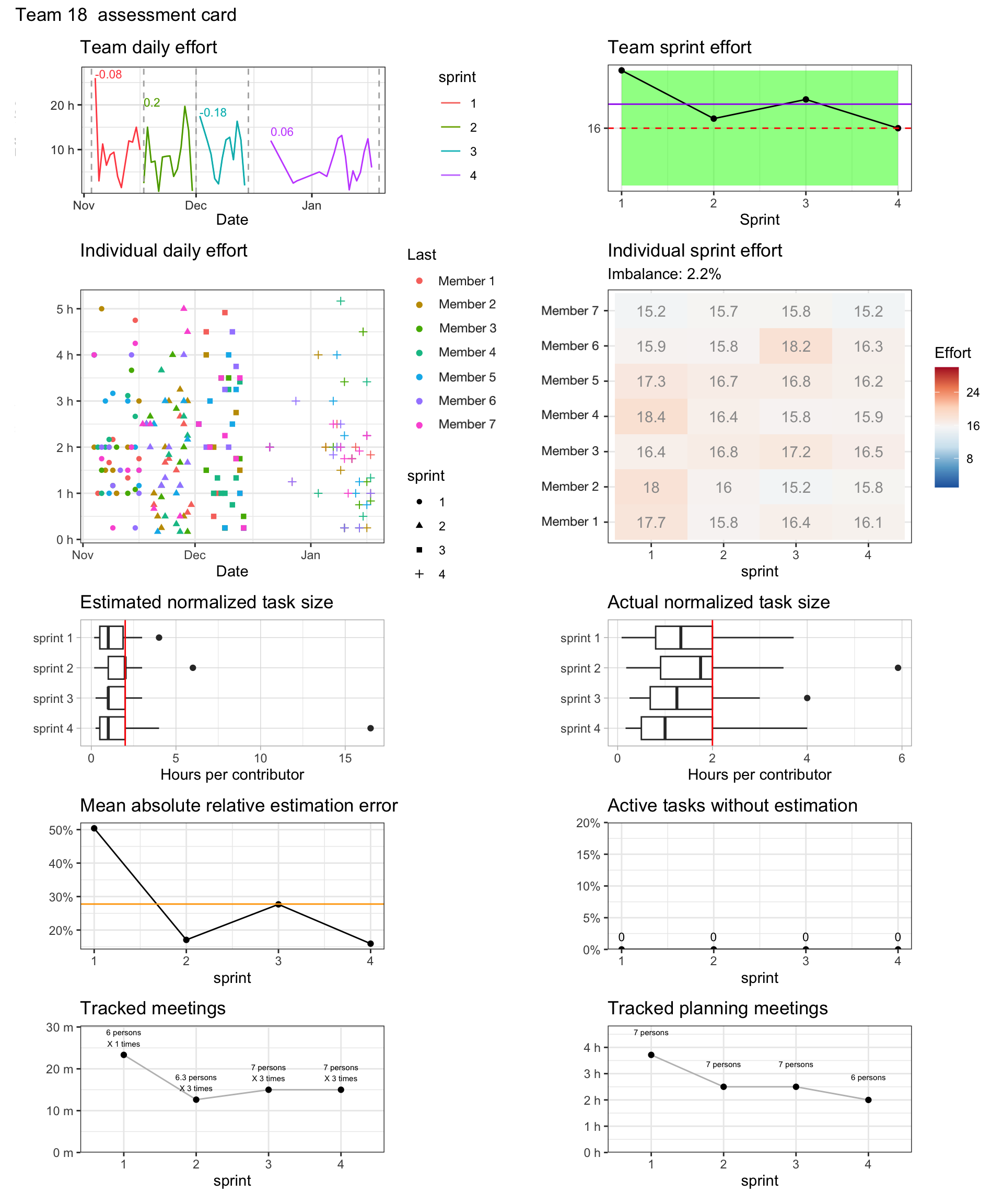}
    \caption{Sample of Team assessment card.}
    \label{fig:team-card}
\end{figure*}

Teams are evaluated using a combination of qualitative and quantitative techniques.
The quantitative assessment is performed through a team assessment card, an example is reported in figure \ref{fig:team-card}).
Qualitative assessment is carried on on the basis of notes taken during project reviews and sprint retrospectives.

\subsubsection{Coordination and improvement}

The criteria we used to assess the team coordination are: 

\begin{itemize}
    \item Compliance with the assigned time budget.
          Looking at the "Team sprint effort" (top right of the card) we check if the effort is close to the time budget $16 \pm 1$ hour per sprint per person. We focus on the latest three sprints and the overall trend.
    
          In the sample card, we observe that the effort is within the acceptable range and getting closer to the target.
             
    \item Effort balanced among members.
          We used the "Individual sprint effort" heat map (second from the top on the right) to detect anomalies. Moreover, we compute the overall imbalance using the Gini index, where good values are below 3\% and bad values are above 9\%.
            
          In the sample card, we appreciate visually a good balance that is confirmed by the Gini index of 2.2\%.
            
    \item Uniform distribution of effort during the sprint.
          The "Team daily effort" line chart (top left of the card) shows the trend of effort, including a per sprint Pearson regression coefficient.
          We can spot strong upward behaviours visually, moreover, we consider r values larger than 0.3 as suspect, and above 0.5 as critical.

          In the sample card, we can see a healthy behaviour with r values below the warning thresholds.
          
    \item Small task size.
        The "Estimated normalized task size" and "Actual normalized task size" boxplots (third row in the card) report the distribution of task size normalized per number of participants. The diagrams feature a reference line corresponding 2 hours per participant which we recommended as the upper limit during the course.
        In general, we consider a good result when teams manage to keep three-quarters of the task below the two-hour threshold. We mainly look at estimation but we also check whether the actual values are kept under control.

        In the sample card, we observe a very good size of tasks.
    
    \item Accurate estimation error. 
          The "Mean absolute relative estimation error" (MRAEE) diagram (second from bottom on the left column) reports the error in the four sprints and the overall average with the orange line.
          In general, we deem excellent MRAEE below 20\% and consider critical an error above 50\%, especially in the latest two sprints.

          In the sample card we observe an initial large estimation error that has been reduced during the project. With an overall average below 30\%.
    
    \item Complete estimation of tasks.
          The "Active tasks without estimation" (second from bottom on the right column) reports the percentage of active tasks -- i.e. tasks with reported effort -- that have no estimation.
          We usually tolerate a very small number of non-estimated tasks, while in general, we consider critical more than 5\% of non-estimated tasks.

          In the sample card, we observe the ideal condition of all tasks being estimated.

    \item Tracked planning and scrum meetings.
         The diagrams "Tracked meetings" and "Tracked planning meetings" (bottom two diagrams in the card) show the average time of scrum and planning meetings respectively. The diagrams also report the number of people involved and the number of occurrences.
         We check, first of all, that meetings are effectively tracked, and also that possibly all members of the team participate. We consider it acceptable that one member is missing in one meeting considering possible personal issues during the semester.

         In the sample card, we observe scrum meetings getting shorter and more frequent -- which is a positive fact -- and one person missing from the latest planning meeting -- which is still acceptable --.
\end{itemize}

We assessed improvement considering the trend along the four sprints of the above criteria.

\subsubsection{Quality of meetings}

We took personal notes during the product review and sprint retrospective meetings that were used to provide immediate feedback to the teams and later to assess the ability of the teams.

\begin{itemize}
    \item Concerning the product reviews, we evaluated the capability to clearly explain which user stories were implemented, to demonstrate them properly on the application, the promptness in the responses to comments from stakeholders, the clarity in the description of the product, and the ability in self-evaluation.
    \item Concerning the retrospective, we evaluated the ability to analyze the team operation during the past sprint, recognize the problems and mistakes committed, and eventually propose sensible improvement actions.
\end{itemize}

\subsubsection{Doneness}

Based on our Definition of Done, we assessed the level of \textit{doneness} of the products delivered at the end of the four sprints, checking whether: the teams performed enough unit and E2E testing and evidence is found in the GitHub repository (e.g., source code of tests, test plans, coverage reports); no software failures appeared during sprint reviews (with a tolerance of one occurrence); TD was managed, i.e. there are explicit tasks on TD reduction that are consistent with the SonarCloud web reports; a running docker image was available in sprints 2 and 4. 

\subsection{Assessment of Team Members}

We evaluated the individual contribution using both quantitative information -- based on the team assessment card -- and qualitative information based on the notes taken during the meetings and the evidence in the repositories. The main criteria we used are:

\begin{itemize}
    \item Active participation in meetings: we assessed this aspect using our notes about how often the student spoke during the meetings, we asked each member to present at least once;
    \item Effective involvement in teamwork: based on the participation we assessed whether the student appeared to be knowledgeable of the overall teamwork, and double-checked it with a few questions on performed activities;
    \item No outliers in daily personal effort: looking at the "Individual daily effort" diagram in the assessment card we checked that the student did not report any abnormal daily effort (e.g. 14 hours in a single day);
    \item Sprint effort within acceptable range: based on the "Individual sprint effort" diagram we checked whether the student reported an effort too far from the individual sprint budget (i.e. 16 hours over two weeks).    
\end{itemize}

\section{Lessons learned}

The five editions of the course that were run allowed us to sum up some lessons learned about the students' work organization and the way to track and assess it:

\begin{itemize}

\item A typical struggle for students is complying with the time budget. A relevant amount of errors in coordination are related to students' time tracking. A typical correction that is applied by students is the \emph{bulk} addition of allocation to tasks at the end of a sprint, in place of rigorous time tracking for all activities performed during the week.

\item By analyzing the improvement in coordination, we acknowledge that most of the teams learn how to self-organize the work by the second sprint. This aspect can be seen as evidence of the feasibility of successfully applying the Scrum framework on a course project, but with an inevitable learning time that includes the duration of the pilot project with workshops (about 2 weeks) and the first sprint (2 weeks) of the project work, so overall one month.

\item We observe that teams tend to self-organize themselves in two sub-teams, dedicated respectively to the back-end and the front-end. This division proves to be not successful since if any of the sub-team is slower in performing tasks, entire sets of stories are not done by the team.

\item Although the effort is fixed, a few teams suffer from self-generated pressure of high productivity, which results in many stories developed without conforming to the definition of done. Usually after sprint 2 they are able to better manage this aspect.

\item The mandatory time tracking activity made the effort balanced between members in most groups, with very few slackers. This aspect is positive in avoiding hostilities among team members.

\item In agreement with the DevX framework~\cite{DevXCACM} we observed that teams that managed to keep task size small -- thus shortening feedback cycles -- were able to achieve better results.




\end{itemize}

\section{Conclusions and actionable tips}

The conduction of five editions of the course allowed us to positively assess the feasibility of teaching Scrum to Master's students with effectiveness in learning and positive teamwork and cooperation. We identify a few main actionable tips for educators aiming at setting up Scrum projects for their students:

\begin{itemize}

\item Do not set a target on scope but only on time (and cost). This aspect proved to be beneficial to avoid teams rushing to complete all the requirements thus paying not enough attention to quality-related activities;

\item Provide a specific time frame for the course project. This aspect proved to be beneficial to better engage and motivate the students, and avoid situations where entire teams were unable to pass the course because of less-producing members.

\item Balance teams according to a variety of factors (skill set, country of origin, gender) to maximize diversity within groups. This aspect usually pays back and prevents dropouts or very low individual performances.

\item In case of team coordination problems, the Scrum master should intervene as soon as possible, to avoid recrudescence of problems and unmanageable/unrecoverable situations.

\end{itemize}

For future editions of the course, our plan is to introduce acceptance criteria specific to the main user stories to have a finer grain control on the delivered projects. We also plan to add universal and technology-agnostic automated End-to-end test cases to verify them. Finally, we will aim at hosting a yearly industrial seminar on how Scrum is applied in a company in the daily practices.  


\bibliographystyle{ACM-Reference-Format}
\bibliography{biblio}

\end{document}